\documentclass[%
 reprint,
 amsmath,amssymb,
 aps,
]{revtex4-1}

\usepackage{hyperref}       % hyperlinks
\hypersetup{
    colorlinks=true,
    linkcolor=blue,
    filecolor=magenta,      
    urlcolor=cyan,
}

\usepackage{graphicx}% Include figure files
\usepackage{dcolumn}% Align table columns on decimal point
\usepackage{bm}% bold math

\graphicspath{{./Figs/}}

\usepackage[utf8]{inputenc}  % needed if you have ä, ü, ... in the source file
\usepackage{csvsimple}

\begin{document}
\preprint{APS/123-QED}

\title{High-Resolution Quantum Cascade Laser Dual-Comb Spectroscopy in the Mid-Infrared with Absolute Frequency Referencing }
% \title{High-Resolution and High-Accuracy Mid-Infrared Spectroscopy with Quantum Cascade Laser Frequency Combs}
% \title{Fast and Accurate High-Resolution Mid-Infrared Spectroscopy with Quantum Cascade Laser Frequency Combs}

\author{K.~N.~Komagata$^1$}
\author{V.~J.~Wittwer$^1$}
\author{T.~Südmeyer$^1$}
\author{L.~Emmenegger$^2$}
\author{M.~Gianella$^2$}
\affiliation{$^1$Laboratoire Temps-Fréquence, Institut de Physique, Université de Neuchâtel, Avenue de Bellevaux 51, 2000 Neuchâtel, Switzerland\\
$^2$Laboratory for Air Pollution / Environmental Technology, Empa, Überlandstrasse 129, 8600 Dübendorf, Switzerland}
\date{\today}

\maketitle

\section{Abstract}
\textbf{ Quantum cascade laser (QCL) frequency combs~\cite{hugi2012midinfrared} have revolutionized mid-infrared (MIR) spectroscopy by their high brightness and fast temporal resolution~\cite{klocke2018singleshot}, and are a promising technology for fully-integrated and cost-effective  sensors~\cite{schwarz2014monolithically}. 
As for other integrated comb sources such as micro-combs~\cite{kippenberg2018dissipative} and interband cascade laser~\cite{bagheri218passively}, QCLs have a comb spacing of several GHz, which is adequate for measurements of wide absorbing structures, typically found in liquid or solid samples~\cite{klocke2018singleshot}.
However, high-resolution gas-phase spectra require spectral interleaving and frequency calibration~\cite{kuse2021frequencyscanned,gianella2022frequency,lepere2022midinfrared}.
We developed a frequency calibration scheme for fast interleaved measurements with combs featuring multi-GHz spacing.
We then demonstrate dual-comb spectroscopy with 600~kHz accuracy in single-shot 54-ms measurements over 40~cm$^{-1}$ using two QCLs at 7.8~$\mathrm{\mu}$m. 
This work is an important contribution towards fast fingerprinting of complex molecular mixtures in the MIR~\cite{gianella2020highresolution}. Moreover, the calibration scheme could be used with micro-combs for spectroscopy and ranging, both in comb-swept~\cite{kuse2021frequencyscanned,riemensberger2020massively} and comb-calibrated setups~\cite{delhaye2009frequency}.
}

% \section{Introduction}
\section{Main}
Over the last two decades, dual-comb spectroscopy~\cite{coddington2016dualcomb} has developed into a powerful tool for applications requiring fast temporal resolution, high spectral accuracy, and broad spectral coverage. It has enabled difficult or previously impossible measurements, such as multi-species trace gas detection in open-path~\cite{giorgetta2021openpath}, monitoring of irreversible processes with micro-second resolution~\cite{klocke2018singleshot}, hyperspectral 3D imaging~\cite{vicentini2021dualcomb}, and fast determination of isotope ratios of multiple species~\cite{parriaux2022isotope}. Moreover, photonic chip integrated comb sources such as micro-combs~\cite{kippenberg2018dissipative}, quantum cascade lasers (QCL)~\cite{hugi2012midinfrared}, and interband cascade lasers~\cite{bagheri218passively}, are potential game-changers, by enabling mass-producible integrated gas sensors for demanding fields, such as health, security, environment, and industrial process monitoring.

However, these integrated comb sources, due to their small size, have large (compared to fiber mode-locked lasers) repetition rates, $f_\mathrm{rep}$, on the order of 10 GHz or higher. The large $f_\mathrm{rep}$, while allowing a high temporal resolution, leads to a sparse spectrum, where narrow molecular absorption are potentially lost between the comb teeth. 

This limitation is not unique to dual-comb spectroscopy or to integrated comb sources, and has been addressed by interleaving many (up to many thousands) offset spectra, where $f_\mathrm{rep}$ and/or the offset frequency, $f_\mathrm{o}$, are tuned in steps~\cite{villares2014dualcomb, muraviev2020broadband, luo2020longwave, hjalten2021optical,lepere2022midinfrared} or continuously~\cite{gianella2020highresolution, kuse2021frequencyscanned}. 

Step-tuning of referenced combs typically consists in stepping the repetition rate by 10s of Hz with dwell times on the order of seconds or more adapted to the measurement. Thus, the full interleaved spectrum is acquired in 10s or 100s of seconds and the frequency accuracy is that of the static comb~\cite{muraviev2020broadband,hjalten2021optical}. 
We note that step-tuning was also used with free running mid-infrared (MIR) QCLs, where a full spectrum was obtained in 1200~s with an accuracy better than 12 MHz~\cite{lepere2022midinfrared}.
Continuous sweeps yield faster measurements with higher spectral point density. However, they require an adapted frequency calibration scheme. 
The endless frequency comb was proposed by Benkler \textit{et al.}~\cite{benkler2013endless}, but it is not convenient in the MIR as it requires electro-optic components and pulsed emission. 
Alternatively, the use of an unbalanced Mach-Zehnder interferometer provided frequency scales with 10-MHz level precision for NIR micro-combs~\cite{kuse2021frequencyscanned} and MIR QCLs~\cite{gianella2022frequency}, with sweep times of 50~$\mathrm{\mu s}$ and 30~ms respectively.

In this work, we present a method to calibrate fast continuous sweeps of multi-GHz frequency combs with sub-MHz accuracy, which can be applied independently of the wavelength range.
We thus demonstrate a QCL-based MIR dual comb spectrometer with a unique combination of fast acquisition, high frequency accuracy, and high spectral point density. 
 
As a proof of concept, we measure 43 transitions of the $\nu_1$ ro-vibrational band of $\mathrm{N_2O}$ and 64 transitions belonging to the $\nu_4$ band of $\mathrm{CH_4}$. 
With the former measurement, we characterize the accuracy of the computed frequency scale and assess the precision of the measurement at varying scanning speeds. The latter measurement serves to improve the literature accuracy of the transition frequencies of $\mathrm{CH_4}$, independently of two recent works~\cite{lepere2022midinfrared,germann2022methane}. 

Rapid scanning and high accuracy are critical features for fingerprinting of complex molecular mixtures and parallel frequency modulated lidar~\cite{riemensberger2020massively}. Accordingly, we trust that our enhanced method for frequency axis calibration is a highly valuable asset for ranging and MIR sensing.

% \section{Results}
% \subsection{Frequency-accurate dual-QCL-comb spectrometer}
\begin{figure*}
    \centering
    \includegraphics[width=\linewidth]{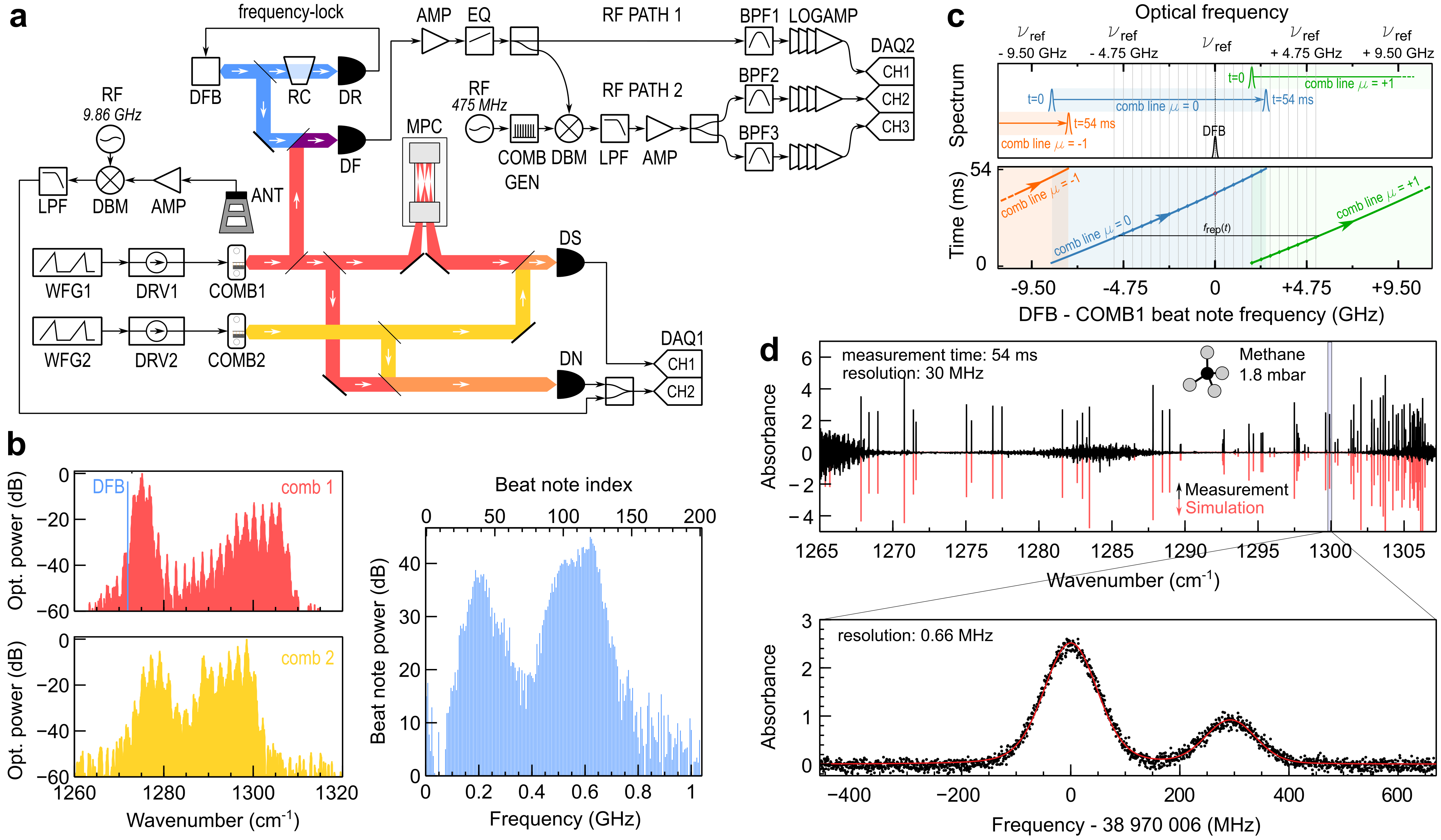}
    \caption{\textbf{Experimental setup and schematic for accurate frequency scale.} (a) Experimental setup featuring the distributed feedback (\texttt{DFB}) laser serving as frequency reference, the gapless dual comb spectrometer, and the setups to measure the repetition rate and the optical frequency during an acquisition. \texttt{MPC}: multi-pass cell, \texttt{AMP}: amplifier, \texttt{BPF}, \texttt{LPF}: band-pass (low-pass) filter, \texttt{DBM}: double-balanced mixer, \texttt{LOGAMP}: logarithmic amplifier, \texttt{EQ}: equalizer, \texttt{COMB GEN}: comb generator, \texttt{WFG}: waveform generator, \texttt{DAQ}: data acquisition unit, \texttt{DS, DN, DR, DF}: photodetector (see main text). (b) Optical spectra of \texttt{COMB1}, \texttt{COMB2}, the position of the reference DFB laser, and the multi-heterodyne beat spectrum. (c) Schematic of the referencing scheme of the optical frequencies of \texttt{COMB1} while they are tuned across more than one free-spectral range ($f_\mathrm{rep}$), relying on the use of a synthetic comb. (d) Absorption spectrum of low pressure methane acquired in $54\,\mathrm{ms}$ by our spectrometer. The top panel is digitally filtered for visibility while the lower panel features two absorption lines at the original resolution of $660\,\mathrm{kHz}$.}
    \label{fig:setup}
\end{figure*}
Our dual-comb spectrometer is composed of two QCL frequency combs~\cite{gianella2020highresolution} (\texttt{COMB1} and \texttt{COMB2}) is schematized in Fig.~\ref{fig:setup}(a). 
\texttt{COMB1} is the interrogating comb which probes the sample contained in a multi-pass cell, and \texttt{COMB2} is the local oscillator.
The repetition frequency of the combs, $f_\mathrm{rep}\approx9.88\,\mathrm{GHz}$, is too large to sample Doppler-broadened gas-phase spectra of small molecules.
Therefore, we tune the driving current to scan the comb lines across $f_\mathrm{rep}$ and interleave spectra to reduce the spectral point spacing by 4 orders of magnitude.
The local oscillator comb is tuned synchronously to keep the multi-heterodyne spectrum within the detection bandwidth of the system [see Fig.~\ref{fig:setup}(b)].
The interferograms measured on the photodetectors \texttt{DS} and \texttt{DN} are divided into \emph{slices} of length $\tau=13.1\,\mathrm{\mu s}$, which are sufficiently long to resolve the beat notes (beat note spacing, $\Delta f_\mathrm{rep}\approx 4.5\,\mathrm{MHz}$).
A lower limit for the spectral resolution, $\delta\nu=\dot{\nu}\cdot(\tau/3.77)$, can be estimated from the slice length, $\tau$, and the tuning rate, $\dot{\nu}$. 
The factor $3.77$ stems from the employed flattop apodization.
For the tuning rates (scan duration) $\dot{\nu}=380\,\mathrm{GHz/s}$~$(27\,\mathrm{ms})$,  $190\,\mathrm{GHz/s}$~$(54\,\mathrm{ms})$, $95\,\mathrm{GHz/s}$~$(107\,\mathrm{ms})$, and $47\,\mathrm{GHz/s}$~$(215\,\mathrm{ms})$ (see below) we find, respectively, $\delta\nu=1.3\,\mathrm{MHz},\,660\,\mathrm{kHz},\,330\,\mathrm{kHz},\,160\,\mathrm{kHz}$.
For example, the spectrometer can acquire spectra spanning $40\,\mathrm{cm^{-1}}$ in $54\,\mathrm{ms}$ at a point spacing and resolution of $0.66\,\mathrm{MHz}$, allowing measurement of low-pressure methane [see Fig.~\ref{fig:setup}(d)]. 
The noise-equivalent absorbance and noise-equivalent absorption coefficient depend on the signal-to-noise ratio (SNR) of the corresponding beat note [see Fig.~\ref{fig:setup}(b)]. 
For strong beat notes, they are of the order of $3\times10^{-4}\,\mathrm{Hz^{-1/2}}$ and $8\times10^{-8}\,\mathrm{cm^{-1}Hz^{-1/2}}$, respectively, given the absorption path length of $36\,\mathrm{m}$ in the multi-pass cell. 
 % Note about NEA: The determination of one absorbance value in the absorbance spectrum requires a measurement of length = 1 slice = 13.1 us. Since we use a flattop apodization, the effective slice length is shortened by a factor of 3.77. The ENBW for a single Fourier coefficient is then 3.77/(2*13.1 us) = 144 kHz. Then, the NEA is (absorbance rms)/sqrt(ENBW).
 
The distributed feedback QCL (\texttt{DFB}) is locked to a molecular transition ($\mathrm{N_2O}$, fundamental $\nu_1$ band, P(14)) and acts as optical frequency reference with frequency $\nu_\mathrm{ref}$~\cite{komagata2022absolute}. 
The beat frequency, $f_b$, between the reference laser and \texttt{COMB1} is detected on a $1.2\,\mathrm{GHz}-$bandwidth detector (\texttt{DF}).
The frequency, $\nu_\mu(t)$, of all lines, $\mu=0,\pm 1,\pm 2,...$, of the interrogating comb over the time of the scan is computed from the measured repetition rate, $f_\mathrm{rep}(t)$, and beat frequency, $f_b(t)$, at each instant of the sweep, and is given by 
\begin{equation}
    \nu_\mu(t)=\nu_\mathrm{ref}+f_b(t)+\mu\cdot f_\mathrm{rep}(t).
    \label{eq:freq-axis}
\end{equation}

\begin{figure*}
    \centering
    \includegraphics[width=\linewidth]{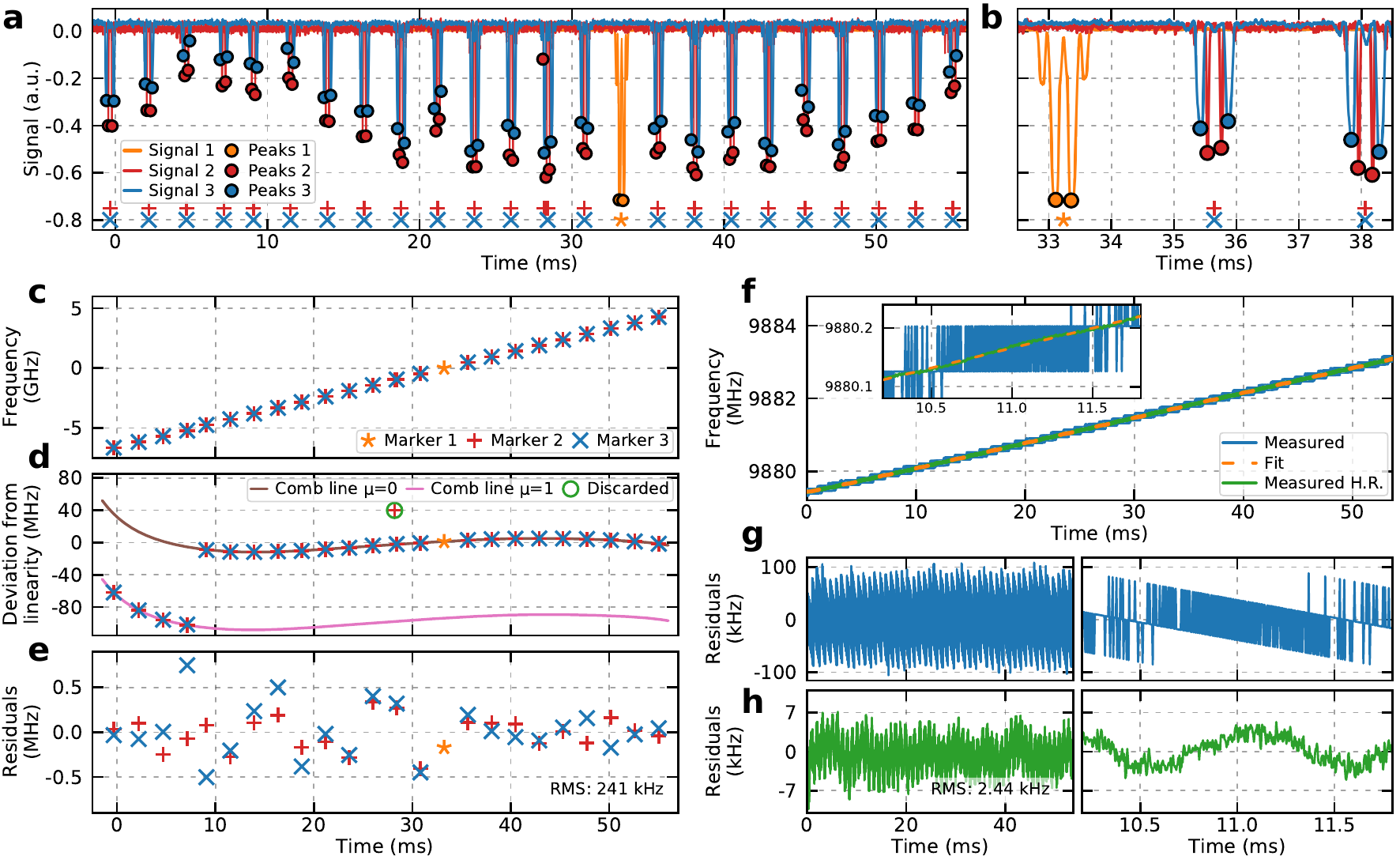}
    \caption{\textbf{Assessment of $f_b$ and $f_\mathrm{rep}$ over a gapless measurement (sweep range $\geq f_\mathrm{rep}$).} (a) Filtered marker signal as a function of time (solid lines) and peaks localized by the algorithm (filled circles). The mean position of two peaks gives the marker position shown by a star, plus or x. (b) Zoom over 3 pairs of peaks. (c) Frequency of the markers as a function of time. (d) Frequency of the markers minus the linear trend and polynomial fitting of the comb line frequency. The first 4 pairs of marker originate from the beat with another comb line ($\mu=1$ instead of $\mu=0$). (e) Residuals from the fit from (d). (f) Measured $f_\mathrm{rep}$ using two methods as a function of time and fit. The inset shows the a zoom over a smaller time interval. (g,h) Residuals of $f_\mathrm{rep}$ measured through the fast processor (g), which is limited by the integration time, or measured with high resolution (H.R.) (h). The right plot shows a zoom over a smaller interval. The fit using the processor data is accurate for the H.R. data within 2.4 kHz. }
    \label{fig:marker}
\end{figure*}

% \subsection{Evaluation of the frequency axis}
The measurement of $f_b(t)$ is based on the open-loop solution proposed in 2009 by Del'Haye \textit{et al.}~\cite{delhaye2009frequency} with a few notable changes. 
Here, the reference laser is fixed while the comb is swept by approximately $f_\mathrm{rep}$. 
Therefore, a single pair of peaks is detected through RF path 1 [see Fig.~\ref{fig:setup}(a)]. 
These are observed at the instants when the band-pass filter \texttt{BPF1} transmits the beat to the rectifying logarithmic amplifier (\texttt{LOGAMP}).
The recorded signal on channel 1 of \texttt{DAQ2} is shown in Fig.~\ref{fig:marker}(a,b) as \emph{Signal 1}.

We implement a second RF path [RF path 2, see Fig.~\ref{fig:setup}(a)], where we mix the beat with a $f_\mathrm{RF}=475\,\mathrm{MHz}$-spaced harmonic RF comb. 
The RF comb provides a synthetic grid of frequencies centered on the reference laser frequency, $\nu_\mathrm{ref}$, and having a spacing of $f_\mathrm{RF}$ [see Fig.~\ref{fig:setup}(c)], whereby the time of passage of $\nu_0$ across each grid line is measured. 
Hence, the determination of $f_b(t)$ (or $\nu_0(t)$) is based on the measurement of the times at which $f_b$ equals multiples of $f_\mathrm{RF}$. 
These markers allow to map the frequency of $f_b(t)$ (or $\nu_0(t)$) versus time, as shown in Fig.~\ref{fig:marker}(c). 
The marker on channel 1 allows identification of $\nu_0(t_0)$ modulo $f_\mathrm{rep}$.

The $5\,\mathrm{GHz}$ RF bandwidth required to measure $f_b$ throughout the sweep, corresponding to $f_\mathrm{rep}/2$, poses a challenge, as the specified bandwidth of the photodetector is only $1.2\,\mathrm{GHz}$. 
We thus used logarithmic amplifiers to increase the marker detection dynamic range and gain equalizers (\texttt{EQ}) to reduce signal strength imbalances.
Moreover, in order to further balance the signal strength and optimize the RF mixers, the signals were split below and above $3.8\,\mathrm{GHz}$ [not shown in Fig.~\ref{fig:setup}(c)]. Further information can be found in Methods and Supplementary Note 1.

We plot the markers after removal of the linear trend in Fig.~\ref{fig:marker}(d). The first 4 sets of markers originate from the beating with a higher comb line $\mu=1$ [see Fig.~\ref{fig:setup}(c)], which was unavoidable to accommodate a sweep about 10 \% larger than $f_\mathrm{rep}$. 
We fit the markers with a 10th order polynomial using a two comb line model separated by $f_\mathrm{rep}(t)$. 
The algorithm is robust against spurious peaks, which produce markers that can be discarded if they don't fit within a certain tolerance region. 
The residuals of the fit are shown in Fig.~\ref{fig:marker}(e). The standard deviation of the fit residuals is found to be $241\,\mathrm{kHz}$.

The repetition rate, detected by the horn antenna, is down-mixed to $\sim20\,\mathrm{MHz}$, digitized on \texttt{DAQ1}, channel 2, and processed together with the interferogram [see Fig.~\ref{fig:setup}(a)].
The resolution is limited to $76\,\mathrm{kHz}$ due to the finite length ($\tau=13.1\,\mathrm{\mu s}$) of the slices [see Fig.~\ref{fig:marker}(f)]. 
%Longer (shorter) interferogram slices would lead to better (worse) frequency resolution and worse (better) temporal resolution in the measurement of $f_\mathrm{rep}$. Clearly, a compromise has to be chosen.
We fit these data using a 10th order polynomial. 
The residuals are limited by the resolution as shown in Fig.~\ref{fig:marker}(g,h). 
On a control measurement, we recorded the raw digitized data from \texttt{DAQ1}, channel 2, and computed $f_\mathrm{rep}(t)$ with higher resolution. 
The residuals between the fit from the low resolution data and the high resolution data are shown in Fig.~\ref{fig:marker}(h), showing correspondence within $2.4\,\mathrm{kHz}$. 
The high-resolution data feature oscillations of $f_\mathrm{rep}$ due to back-reflection from the photodetectors.

% Retrieved line center accuracy ================================================================
% \subsection{Retrieved line center accuracy}
\begin{figure*}
    \centering
    \includegraphics[width=\linewidth]{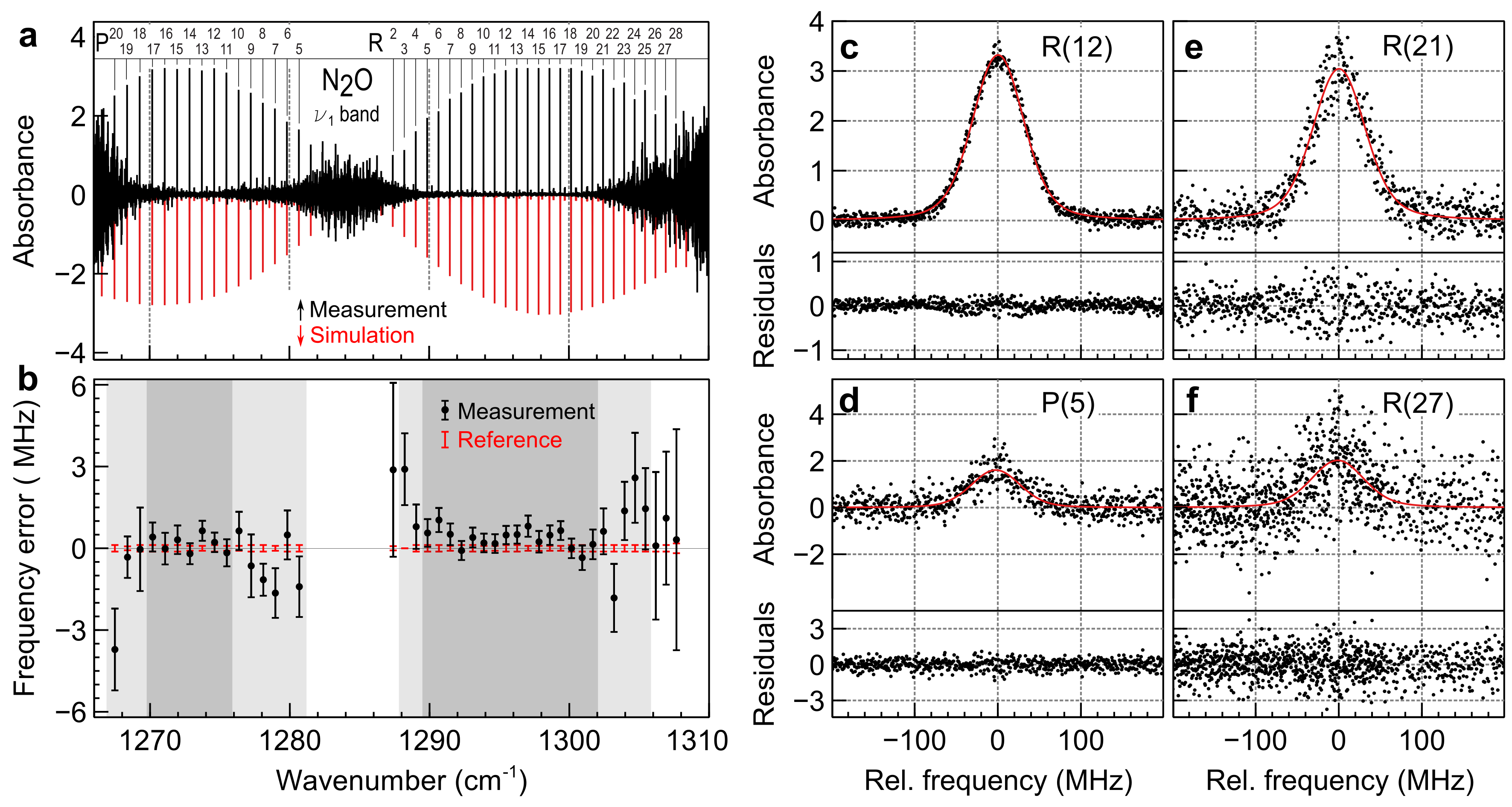}
    \caption{\textbf{Spectra and retrieved line centers in a $54\,\mathrm{ms}$ acquisition} (a) Single shot spectra of the $\nu_1$ fundamental band of nitrous oxide with $660\,\mathrm{kHz}$ spectral point spacing, acquired in 54-ms and comparison with simulation. (b) Retrieved line centers compared with literature reference with 1-$\sigma$ error bars. (c-f) Zoom of (e) on 4 different lines illustrating stronger/weaker absorbance and high/low SNR, and residuals of the Voigt fit shown in red.}
    \label{fig:spectrum}
\end{figure*}

Figure~\ref{fig:spectrum}(a) shows a spectrum of the fundamental $\nu_1$ band of nitrous oxide acquired in a single sweep of $54\,\mathrm{ms}$.
The bimodal comb amplitude distribution gives rise to a variable SNR across the spectrum, as can be seen in Figs.~\ref{fig:spectrum}(c-f). We used a Voigt fit to retrieve the line centers of the molecular transitions.

We benchmark the frequency accuracy of our spectrometer by comparing the computed line center frequencies to those obtained by two other groups with 100-kHz level accuracy~\cite{alsaif2018high, hjalten2021optical}.
Based on multiple sets of measurements over the course of a few weeks, we estimate that we retrieve, in a single-shot acquisition, 40 line centers (spanning $30.4\,\mathrm{cm^{-1}}$) with a 1-$\sigma$ accuracy better than $2\,\mathrm{MHz}$, among which 23 lines (spanning $17.1\,\mathrm{cm^{-1}}$) are retrieved with an accuracy better than $0.6\,\mathrm{MHz}$, as can be seen in Fig.~\ref{fig:spectrum}(b).

\begin{figure*}
    \centering
    \includegraphics{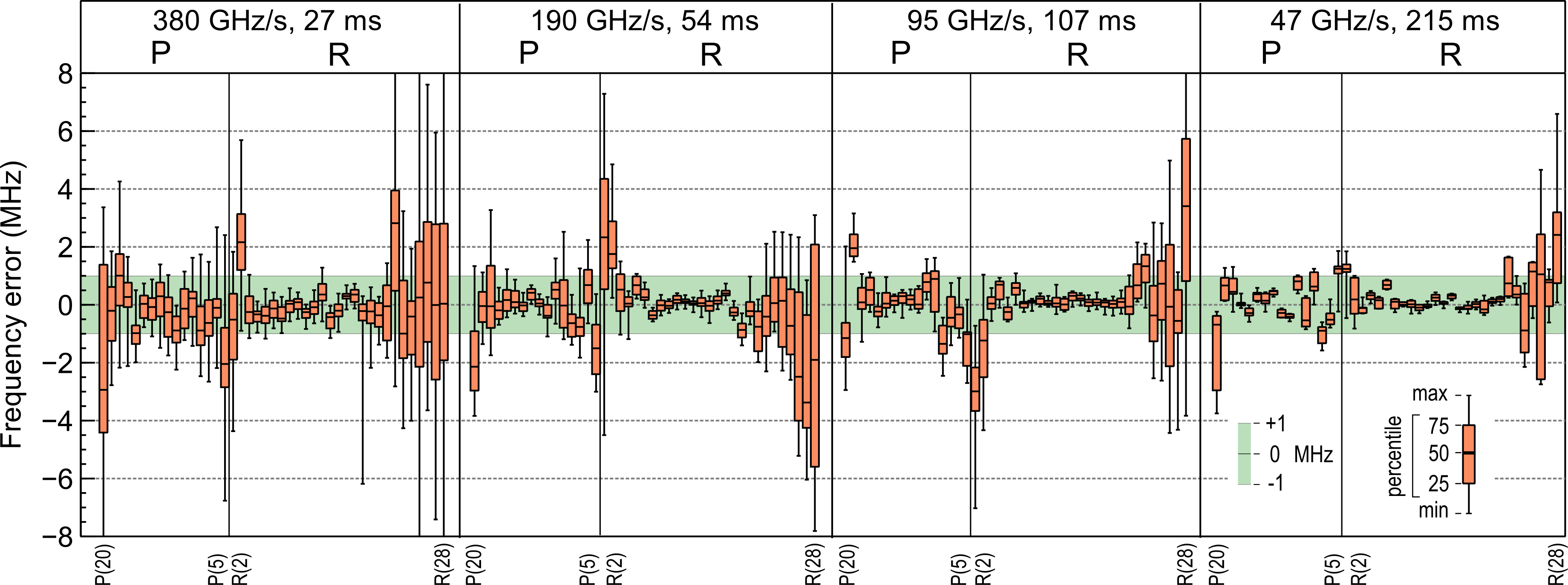}
    \caption{\textbf{Frequency precision for varying scanning speed.} Frequency error and statistical spread of the retrieved line centers at 4 different scan speeds, decreasing from left to right by factors of 2. The precision increases with the scan time.}
    \label{fig:scanning_speed}
\end{figure*}

This estimation includes the uncertainties of the reference laser frequency $\nu_\mathrm{ref}$ ($180\,\mathrm{kHz}$ after  correction of a  systematic bias of $(180\pm100)\,\mathrm{kHz}$), $f_\mathrm{rep}$ ($2.6\,\mathrm{kHz}$ after the correction of a systematic bias of $(2.9\pm0.7)\,\mathrm{kHz}$), the fit of the markers as in Fig.~\ref{fig:marker}(e) ($240\,\mathrm{kHz}$), and the standard deviation of the retrieved line center frequencies over 12 consecutive 54-ms measurements (below $300\,\mathrm{kHz}$ for the lines with the best SNR, see Fig.~\ref{fig:scanning_speed}).
The retrieved line centers and the uncertainties are consistent with the literature values.

The frequency accuracy of the spectrometer allows us to retrieve the line centers of $\mathrm{CH_4}$ with a higher accuracy than the current literature~\cite{gordon2022hitran2020}. Our results are given in Table~\ref{table:methane}. For this measurement, we averaged the retrieved line centers over 15 measurements.
\begin{table}
\centering
\begin{tabular}{|c|c|}\hline%
\bfseries Transition & \bfseries Line center\\
(J',C',$\alpha$') - (J'',C'',$\alpha$'') & (cm$^{-1}$)\\\hline
\begin{filecontents*}{transitionExport2.csv}
Transition;FrequencyMeas
( 5, A2, 2) - ( 6, A1, 1);1267.822296(55)
( 5, F2, 4) - ( 6, F1, 1);1268.367774(65)
( 5, F1, 3) - ( 6, F2, 2);1268.976244(54)
( 5, A1, 1) - ( 6, A2, 1);1270.785068(25)
( 5, F1, 4) - ( 6, F2, 1);1271.406911(12)
( 5,  E, 3) - ( 6,  E, 1);1271.589462(34)
( 4, F2, 3) - ( 5, F1, 2);1275.041674(17)
( 4,  E, 2) - ( 5,  E, 1);1275.386810(24)
( 4, F1, 3) - ( 5, F2, 1);1276.844312(23)
( 4, F2, 4) - ( 5, F1, 1);1277.473279(42)
( 3, F1, 3) - ( 4, F2, 1);1281.610531(41)
( 3,  E, 2) - ( 4,  E, 1);1282.62456(11)
( 3, F2, 2) - ( 4, F1, 1);1282.984142(95)
( 2, A1, 1) - ( 3, A2, 1);1287.813273(40)
( 2, F1, 2) - ( 3, F2, 1);1288.457016(25)
( 2, F2, 2) - ( 3, F1, 1);1288.950985(35)
(10, F1, 3) - (10, F2, 1);1292.610277(23)
(10, A1, 2) - (10, A2, 1);1292.677897(17)
( 1, F1, 1) - ( 2, F2, 1);1294.379266(18)
( 1,  E, 1) - ( 2,  E, 1);1294.684342(20)
( 9, F2, 3) - ( 9, F1, 1);1295.212700(23)
( 9, F1, 3) - ( 9, F2, 1);1295.326105(16)
( 8, A2, 1) - ( 8, A1, 1);1297.486522(14)
( 8, F2, 3) - ( 8, F1, 1);1297.656959(16)
( 8,  E, 2) - ( 8,  E, 1);1297.750892(17)
(11, A1, 2) - (11, A2, 1);1297.819233(21)
(11, F1, 5) - (11, F2, 2);1298.205596(29)
(11,  E, 3) - (11,  E, 1);1298.437452(34)
( 7, F2, 2) - ( 7, F1, 1);1299.629814(14)
( 7, F1, 3) - ( 7, F2, 1);1299.899492(17)
( 6,  E, 2) - ( 6,  E, 1);1301.371150(21)
( 9,  E, 3) - ( 9,  E, 1);1301.491649(35)
( 6, F1, 2) - ( 6, F2, 1);1301.553893(25)
( 9, F2, 4) - ( 9, F1, 2);1301.824890(29)
( 6, A1, 1) - ( 6, A2, 1);1302.044327(19)
( 8, F1, 3) - ( 8, F2, 1);1302.777043(46)
( 5, F2, 2) - ( 5, F1, 1);1302.787450(41)
( 9, A2, 2) - ( 9, A1, 1);1302.950378(34)
(10, A2, 2) - (10, A1, 1);1303.200731(70)
( 5, F1, 2) - ( 5, F2, 1);1303.378292(32)
( 7, A1, 2) - ( 7, A2, 1);1303.571715(52)
( 4, A2, 1) - ( 4, A1, 1);1303.712214(69)
( 8, F2, 4) - ( 8, F1, 2);1303.955066(59)
( 4, F2, 2) - ( 4, F1, 1);1304.232185(53)
( 7, F1, 4) - ( 7, F2, 2);1304.472072(76)
( 4,  E, 1) - ( 4,  E, 1);1304.599775(81)
( 3, F2, 1) - ( 3, F1, 1);1304.857914(99)
( 6, F1, 3) - ( 6, F2, 2);1305.002823(77)
( 7,  E, 2) - ( 7,  E, 1);1305.024853(57)
( 9, F2, 5) - ( 9, F1, 3);1305.106730(88)
( 3, F1, 2) - ( 3, F2, 1);1305.400123(61)
( 2,  E, 1) - ( 2,  E, 1);1305.43730(11)
( 2, F1, 1) - ( 2, F2, 1);1305.707066(98)
( 5, F2, 3) - ( 5, F1, 2);1305.903988(62)
( 4, F1, 2) - ( 4, F2, 1);1305.98730(10)
( 6, A2, 1) - ( 6, A1, 1);1306.269344(86)
\end{filecontents*}
\csvreader[
head to column names,
separator=semicolon,
late after line=\\,
late after last line=\\\hline
]{transitionExport2.csv}{}{%
\Transition & \FrequencyMeas
}%
\end{tabular}
\caption{Measured line list of the methane $\nu_4$ band. The upper (lower) levels are labeled with a single (double) prime and are denoted according to the Hitran~\cite{gordon2022hitran2020} notation.}
\label{table:methane}
\end{table}

% Scanning speed ============================================================
% \subsection{Scanning speed}
The scanning speed of the spectrometer can be varied to measure spectra in between $27\,\mathrm{ms}$ to $216\,\mathrm{ms}$.  
We estimate an accumulated chirp~\cite{riemensberger2020massively} rate of $44\,\mathrm{THz/s}$ to $5.6\,\mathrm{THz/s}$, considering 120 lines which sweep in parallel. 

The sweep becomes discontinuous at faster rates.  
Slower tuning is possible, but requires a larger buffer in the acquisition unit.

Figure~\ref{fig:scanning_speed} displays the statistics in the form of box plots of the retrieved $\mathrm{N_2O}$ line centers with respect to literature reference for varying scanning speeds. 
Scan speeds up to $380\,\mathrm{GHz/s}$ do not induce noticeable offsets in the retrieved line centers, within the uncertainties.
This shows that delays due to the RF scheme are negligible. 
The measurement at 380~GHz/s was performed at a later date. Potential delays at faster scanning speeds could be assessed and corrected from a reference measurement.

% Discussion =====================================================================
% \section{Discussion}
We have demonstrated a fast and accurate mid-infrared dual-comb spectrometer. 
The combs were emitted by quantum cascade lasers, which can be tuned over their free spectral range ($f_\mathrm{rep}$) in tens of milliseconds to provide high-resolution spectra of broad molecular bands. 
We calibrated the frequency axis by measuring the beat between the interrogating comb and a reference laser and by acquiring the comb's repetition rate. 
Moreover, we have enhanced the marker method~\cite{delhaye2009frequency} using a synthetic comb, to process the fast chirp of the beat between the comb and the reference laser. This method is compatible with other integrated comb technologies featuring multi-GHz repetition frequencies such as micro-resonator combs~\cite{kippenberg2018dissipative}, and could also be applied to comb-calibrated spectroscopy~\cite{delhaye2009frequency}.

Thus, our agile spectrometer provides unique performances in terms of acquisition speed and frequency accuracy compared to other broadband spectrometers with small ($<10\,\mathrm{MHz}$) spectral point spacing~\cite{delhaye2009frequency, nishiyama2013high, alsaif2018high, gotti2020comblocked, hjalten2021optical}. 
Our spectrometer is a promising candidate towards an on-chip gas sensor that is highly sensitive and selective to many relevant molecules, allowing real-time monitoring of complex gas mixtures at Hz-level refresh rates.

Finally, we believe that the spectrometer sensitivity, bandwidth, and power-efficiency will improve thanks to ongoing development in quantum cascade laser frequency combs, such as mutual stabilization~\cite{komagata2021coherentlyaveraged, hillbrand2022synchronization}, dispersion management~\cite{villares2016dispersion}, and full spectrometer integration~\cite{schwarz2014monolithically}. 

% Methods ===================================================================
\section{Methods}
\subsection{Experimental setup}
% Lasers
The two laser sources were InGaAs/AlInAs on InP-based dual-stack QCLs~\cite{jouy2017dual,gianella2020highresolution}. They were $4.5\,\mathrm{mm}$ long ($f_\mathrm{rep}\approx9.88\,\mathrm{GHz}$, $\Delta f_\mathrm{rep}\approx4.5\,\mathrm{MHz}$) and had both end facets uncoated. 
The combs emitted several hundred mW of optical power over $\sim50\,\mathrm{cm^{-1}}$ in the $1300\,\mathrm{cm^{-1}}$ range [see Fig.~\ref{fig:setup}(b)].
They were operated at constant temperature of $1.9^{\circ}\mathrm{C}$ and $2.2^{\circ}\mathrm{C}$, respectively.
The combs were electrically driven by a pair of QubeCL drivers (ppqSense).
The bias currents for the two combs were $1532\,\mathrm{mA}$ and $1226\,\mathrm{mA}$, respectively.
The currents were modulated by asymmetric triangular waveforms (\texttt{WFG1} and \texttt{WFG2} are the two outputs of an ArbStudio 1102, LeCroy waveform generator) with the currents initially decreasing at constant rate during a time $T=30\,\mathrm{ms}$, $60\,\mathrm{ms}$, $120\,\mathrm{ms}$, or $240\,\mathrm{ms}$ and returning to the initial value over a shorter time $0.25\cdot T$.
The peak-to-peak current modulation amplitude was approximately $90\,\mathrm{mA}$ for both lasers.
This required a modification of the QubeCL drivers which, by default, do not allow such large current modulations.
The beam splitters were custom-made to obtain the appropriate power ratios at each photodetector. They were designed for reflecting 1\%, 10\% or 50\% at an angle of incidence of 45° for s-polarized light from 1220-1370$\,\mathrm{cm}^{-1}$. They were produced by ion beam sputtering (IBS) on 5 mm thick wedged (0.5 deg) $\mathrm{CaF_2}$ substrates with an anti-reflection coating on the back side to minimize undesired interference effects.

% Detectors, data acquisition
Two $1\,\mathrm{GHz}-$bandwidth MIR photodetectors were employed to measure the sample (\texttt{DS}) and to normalize laser intensity and phase noise (\texttt{DN}).
The detector outputs were sampled at $2.5\,\mathrm{GSa/s}$ with 12 bit resolution by the data acquisition unit \texttt{DAQ1} (ADQ32, Teledyne), generating $10\,\mathrm{Gbyte/s}$ of raw data for the duration of the sweep. 
The data acquisition was triggered at $t=0.07\cdot T$ after the start of the (decreasing) current ramp and continued for a duration of $27\,\mathrm{ms}$, $54\,\mathrm{ms}$, $107\,\mathrm{ms}$, or $215\,\mathrm{ms}$, which corresponds to $2^{26}$, $2^{27}$, $2^{28}$, or $2^{29}$ samples at $2.5\,\mathrm{GSa/s}$.
The clocks of \texttt{DAQ1} (Fig.~\ref{fig:setup}(a)) and of the synthesizer generating the $475\,\mathrm{MHz}$ fundamental for the synthetic comb generator were referenced to a GPS-disciplined clock (GPSDO, Leo Bodnar).

% Data processing
A custom software (C++ and CUDA) was used to process the two interferograms.
The program is capable of fully processing (from raw data to absorbance values) 25 $27\,\mathrm{ms}$-long sweeps per second, corresponding to a raw data throughput of $6.75\,\mathrm{Gbyte/s}$, or 67.5\% of the maximum raw data input rate.
The interferograms were first divided into \emph{slices} of length $2^{15}=32,768$ samples, corresponding to $13.1\,\mathrm{\mu s}$.
The beat note amplitudes were computed for each slice by fast Fourier transform, from which the sample's absorbance was determined.

% Optical reference
A distributed feedback QCL (\texttt{DFB}, Alpes Lasers) emitting a continuous wave (cw) beam was locked to a molecular transition ($\mathrm{N_2O}$, fundamental $\nu_1$ band, P(14)) by means of a wavelength modulation scheme and acted as optical frequency reference with frequency $\nu_\mathrm{ref}$. 

\subsection{Measurement of the markers and retrieval of the frequency axis}
The heterodyne beat between the DFB reference laser and \texttt{COMB1} is detected on the fast photodetector (UHSM-10.6 PV-4TE-10.6-0.05-butt, Vigo System). 
Through RF path 1, two pulses are transmitted in rapid succession when $f_b(t)$ passes near zero and are amplified by a logarithmic amplifier (ZX47-60LN-S+, Minicircuits).
The timings of the two peaks, $t_0^{\pm}$, fulfill $f_b(t_0^{\pm})=\pm f_\mathrm{bp,1}$, where $f_\mathrm{bp,1}=25\,\mathrm{MHz}$ is the center frequency of \texttt{BPF1}.
We assume linear chirping of the beat between $t_0^-$ and $t_0^+$, so that $f_b(t_0)=0$, where $t_0$ is the mean of $t_0^+$ and $t_0^-$.
With Eq.~\ref{eq:freq-axis} we find $\nu_0(t_0)=\nu_\mathrm{ref}$.

As for RF path 2, $f_b(t)$ is mixed with an RF comb with spacing $f_\mathrm{RF}$. Therefore, each time the \emph{DFB-comb} beat, $f_b(t)$, nears a multiple of $f_\mathrm{RF}$, two double pulses are recorded on channels 2 and 3 of \texttt{DAQ2} (\emph{Signal 2} and \emph{Signal 3} in Fig.~\ref{fig:marker}(a,b)).
The timings, $t_n^{\pm}$ of the two pulses of the $n$th ($n=\pm1, \pm2,...$) double pulse fulfill $f_b(t_n^{\pm}) = n\cdot f_\mathrm{RF} \pm f_\mathrm{bp,i}$, where $f_\mathrm{bp,2}=21\,\mathrm{MHz}, f_\mathrm{bp,3}=45\,\mathrm{MHz}$ are the center frequencies of \texttt{BPF2} and \texttt{BPF3}, respectively.
For the mean times, $t_n=(t_n^-+t_n^+)/2$, we find, as before, $f_b(t_n)=n\cdot f_\mathrm{RF}$ and with Eq.~\ref{eq:freq-axis}, $\nu_0(t_n)=\nu_\mathrm{ref}+n\cdot f_\mathrm{RF}$. More details concerning the processing of the markers can be found in Supplementary Figure 1 and Supplementary Note 1.

As post-processing, we first smooth the marker signals with a low-pass filter. 
Then we detect the peaks which match specific criteria in terms of prominence, width, and proximity to other lines using a peak-finding algorithm (find\_peaks, scipy). 
If two peaks have the appropriate time delay, we establish that $f_b$ was at a multiple integer $n$ of $f_\mathrm{RF}$ at the mean time of the two peaks. 
The integer $n$ is guessed from linear interpolation using marker 1 and an estimated chirp. 
The discontinuity from the change of comb line can be easily recognized and the markers originating from comb line $\mu=1$ can be used to determine the frequency of comb line $\mu=0$ by knowledge of $f_\mathrm{rep}(t)$.
%and the corresponding markers are translated, with knowledge of $f_\mathrm{rep}(t)$, within the domain $[-f_\mathrm{RF}/2,f_\mathrm{RF}/2]$

We fit the markers with polynomials of increasing orders up to 10, keeping the coefficients of the previous order as initial guess and providing the initial guess for the new coefficient from the residuals of the previous fit. 
A larger weight is given to marker 2 as \texttt{BPF2} is narrower than the other band-pass filters. 
After the fits with the $4^\mathrm{th}$ and $8^\mathrm{th}$ order polynomial, markers outside a tolerance of $50\,\mathrm{MHz}$ and $10\,\mathrm{MHz}$ are discarded, respectively.
This allows the removal of most spurious peaks, which do not align in the plot Fig.~\ref{fig:marker}(d).

\subsection{Measurement of the repetition rate}
In QCL frequency combs, the beating of all co-existing modes in the laser produces a measurable voltage modulation on the laser electrodes with frequency equal to $f_\mathrm{rep}$~\cite{piccardo2018timedependent}.
The laser module features a $50\,\mathrm{\Omega}$ transmission line on a printed circuit board that connects an SMA connector at the back of the housing to a point very close (within $1\,\mathrm{mm}$) to the back facet of the laser.
From there, a short bond wire establishes an electrical connection to the laser's top electrode.
This pathway can be used for RF injection or to measure $f_\mathrm{rep}$.
However, connecting cables and other circuitry to the SMA port tends to increase laser phase noise. 
It is observable as a broadening of the beat notes.
Therefore, a horn antenna (\texttt{ANT}, PowerLOG 40040, Aaronia) was pointed at the open SMA connector. The detected $f_\mathrm{rep}$ was then amplified and down-mixed to $\sim20\,\mathrm{MHz}$ to enable its measurement on one channel of \texttt{DAQ1}.

\subsection{Gas sample preparation and measurement}
We dynamically diluted $\mathrm{N_2O}$ gas with nitrogen to a concentration of about of $450\,\mathrm{ppmv}$ and flowed the diluted mixture through the multi-pass cell (\texttt{MPC}).
We then closed off the cell and pumped down to $1.5\,\mathrm{mbar}$, which provided sufficiently strong absorption lines and sufficiently low pressure to be reasonably close to the Doppler-broadened limit and, more importantly, to minimize the pressure-induced shift on the measured transitions.
Multiple sweeps were taken but not averaged together.
The time delay between sweeps was several seconds, to ensure proper synchronization of the two data sets acquired by the two DAQs, which were read out by independent software.
By integrating both DAQs in the same processing routine, the acquisition rate (sweeps per unit time) could be increased significantly (about 12 $54\,\mathrm{ms}$-long sweeps per second).

\section*{Acknowledgments}
We thank Prof. Jérôme Faist and Dr. Mathieu Bertrand at ETH for providing the QCL combs and lending the horn antenna used in this work. We also thank Dr. Stéphane Schilt for fruitful discussions concerning the marker scheme.

\section*{References}
\bibliography{main}% Produces the bibliography via BibTeX.

\end{document}